\documentclass[]{spie}  

\usepackage{makecell}
\usepackage[table]{xcolor}
\usepackage{amsmath,amsfonts,amssymb}
\usepackage{graphicx}
\usepackage[colorlinks=true, allcolors=blue]{hyperref}
\usepackage{optidef}
\usepackage{comment}
\usepackage{caption}
\usepackage{subcaption}

\title{Compressed learning based onboard semantic compression for remote sensing platforms}

\author[a]{Protim Bhattacharjee}
\author[a]{Peter Jung}
\affil[a]{Institute of Optical Sensor Systems, German Aerospace Center (DLR), Berlin, Germany}

\authorinfo{Further author information: (Send correspondence to P.B.)\\P.B.: E-mail: protim.bhattacharjee@dlr.de\\  P.J.: E-mail: peter.jung@dlr.de}

\begin{document} 
\maketitle

\begin{abstract}
Earth observation (EO) plays a crucial role in creating and sustaining a resilient and prosperous society that has far reaching consequences for all life and the planet itself. Remote sensing platforms like satellites, airborne platforms, and more recently dones and UAVs are used for EO. They collect large amounts of data and this needs to be downlinked to Earth for further processing and analysis. Bottleneck for such high throughput acquisition is the downlink bandwidth. Data-centric solutions to image compression is required to address this deluge. In this work, semantic compression is studied through a compressed learning framework that utilizes only fast and sparse matrix-vector multiplication to encode the data. Camera noise and a communication channel are the considered sources of distortion. The complete semantic communication pipeline then consists of a learned low-complexity compression matrix that acts on the noisy camera output to generate onboard a vector of observations that is downlinked through a communication channel, processed through an unrolled network and then fed to a deep learning model performing the necessary downstream tasks; image classification is studied. Distortions are compensated by unrolling layers of NA-ALISTA with a wavelet sparsity prior. Decoding is thus a plug-n-play approach designed according to the camera/environment information and downstream task. The deep learning model for the downstream task is jointly fine-tuned with the compression matrix and the unrolled network through the loss function in an end-to-end fashion. It is shown that addition of a recovery loss along with the task dependent losses improves the downstream performance in noisy settings at low compression ratios.  
\end{abstract}


\section{INTRODUCTION}
\label{sec:intro}  
With the advances in the field of computer vision and computing hardware, the need and usage of remote sensing products for Earth observations is increasing rapidly. It is now possible to have near-real-time data for natural disasters, or monitoring vegetation change across seasons, urban mapping  and, among other use-cases, surveillance of critical infrastructures on land and at sea. Various platforms are also used for such remote observations, namely, satellites, unmanned aerial vehicles (UAVs), and drones. Consequently, the data archives of these various missions are growing in size; the Sentinel mission data archives alone already exceed 78 PB~\cite{sentinel_dashboard}  and is increasing. This large data volume must be downlinked from the remote platform. However, the downlink bandwidth for all these platforms is limited and this may cause reasonable delays in communication and data analysis. For example, during a first-responder rescue mission for natural disasters, there may only be a low data rate link possible between the remote platform and the control center. Downlink of large imagery may not possible or be time consuming. This will lead to delays in generation of different remote sensing data products like infrastructure maps or change detection maps. Thus, data-centric compression methods are required to address this deluge. Though compute on remote platforms have increased significantly with the use of components-off-the-shelf (COTS), minimizing resource utilization is still an important design criterion. The pivotal idea is that in most cases the final product is derived from the downlinked data and represents some usecase such as land use and land cover (LULC) or object detection. Semantic compression is a paradigm that aims to downlink only that information that is useful for the downstream task, e.g. classification or segmentation. Encoding-decoding architectures are well studied for this purpose. However, most studies involve performing a deep learning based encoding onboard. However, it may not be possible, especially in case of UAVs and high altitude platforms (HAPs) to implement state-of-the-art encoders, e.g. a vision transformer, due to insufficient memory and compute onboard. In this work, semantic compression is studied from the perspective of compressed learning and a linear compressor is proposed that implements a single matrix-vector multiply to perform the feature encoding. Previously, such a linear compressor is learned either with no consideration of physical constraints or through loss function regularization that enforces physical constraints such as binary matrices~\cite{arguello_binary_matrix}. In this work the linear compressor is described by a categorical distribution that is learned along with the required downstream task using the GLODISMO~\cite{glodismo} framework that has the advantage of directly learning and optimizing for physically plausible structured discrete matrices without additional regularization. The downstream task is performed with a deep learning model is proposed to be implemented at the groundstation. Thus, reducing the required onboard compute. Further camera noise; additive Gaussian, photon noise, and impulse noise and an AWGN communication channel are considered as the source of distortions. The complete semantic compression pipeline then consists of a learned compression matrix that acts on the noisy camera output to generate onboard a vector of observations that is downlinked through an AWGN channel, processed through a unrolled network and then fed to a deep learning model performing the necessary downstream tasks. The semantic nature of the pipeline arises from the fact that we train the compression matrix with the task. It is shown that the proposed framework adds robustness to the downstream task by inclusion of a recovery loss along with the task dependent losses.

Section~\ref{sec:sig_model} describes the mathematical problem considered. Section~\ref{sec:litRev} provides a review of the current state-of-the-art with respect to compressed learning.  The datasets, experiments, and results from the simulations is presented in Sec.~\ref{sec:results}. The article concludes with a discussion and directions of future work in Sec.~\ref{sec:discussion}.  

\section{End-to-end semantic compression signal model}
\label{sec:sig_model}
In remote sensing applications of deep learning the performance of the learned model on the downstream task is of utmost importance. The downstream task can be written as $c = F_u(x)$, where $F_u$ is a neural network model with weight parameters $u$ responsible for generating semantic classes for the scene $x \in \mathbb{R}^N$, $N$ is the total no. of pixels representing the scene, that is observed from the remote platform. Consider the following signal model, 
\begin{equation}
    x = \Psi'\alpha,
\end{equation} $\Psi \in \mathbb{R}^{N \times N}$ is a representation or sparsity basis for the signal and $\alpha \in \mathbb{R}^N$ is the corresponding representation coefficient vector. The linear measurement or compression model can then be written as
\begin{equation}
    y = \Phi(x + n),
\end{equation}
where $\Phi \in \mathbb{R}^{m \times N}$ is the measurement or compression matrix $n \in \mathbb{R}^N$ is the camera noise, and $y \in \mathbb{R}^m$ is the compressed vector of size $m$. The ratio, $R =\frac{m}{N}$ is called the compression ratio. An AWGN channel model is considered,  
\begin{equation}
    z = Hy + w,
\end{equation} where $H \in \mathbb{R}^{m \times m}$ is the channel matrix and $w \in \mathbb{R}^m$ represents the possible quantization noise and the additive channel noise. The onboard platform downlinks $z$. Thus, compression is achieved as generally $z$, respectively $y$, is much smaller than $x$, i.e., $m << N$. For decoding, an equalization may be performed on $z$ to remove the channel and quantization effects to obtain $\hat{y}$, which is then passed to a reconstruction algorithm that solves (\ref{eq:recon}) and corrects for the camera noise. 
\begin{argmini}{\alpha}
{\|\hat{y} - \Phi \Psi' \alpha\|_2^2 + \lambda\|\alpha\|_1.}
{\label{eq:recon}}
{\hat{\alpha} = }
\end{argmini} One can solve~(\ref{eq:recon}) by unrolling the reconstruction algorithm as layers of a neural network~\cite{unrolling_eldar}. Thus, \begin{equation}
    \hat{\alpha} = f_{\theta}(\hat{y}),
\end{equation} where $\theta$ are the learnable parameters of the unrolled network.  The reconstructed signal, $\hat{x}$, can then be obtained as\begin{equation}{\label{eq:syn}}
\hat{x} =   \Psi'\hat{\alpha}
\end{equation} This reconstructed signal is then input to $F$ to evaluate the final class outputs, $\hat{c} = F_u(\hat{x})$. For an end-to-end network learning we have the set $\{ \tilde{\phi}, \tilde{\Psi}, \tilde{\theta}, \tilde{u} \}$ representing the learnable parameters of the compression matrix, representation basis, unrolled reconstruction algorithm, and the neural network weights respectively. As the neural network weights representing the downstream task are learnt along with the compression matrix $\Phi$, semantic compression is achieved by downlinking of the compressed representation of the image $z$ through the channel $H$. The next section discusses techniques and architectures developed in the literature to learn such end-to-end compression networks.

\section{Literature Review}
\label{sec:litRev}

One of the first works to consider prediction from compressed measurement was~\cite{smashedFilter}. It poses the classification as a hypothesis testing problem and provides a framework for generalized maximum likelihood classification. Random binary orthoprojectors were used as compression matrices for generating the compressive measurements. 
In~\cite{lohit_cl} a classification network based on CNNs was used. The compression matrix, $\Phi$ is hand designed for different datasets and is fixed. Gaussian random matrices and permuted Hadamard matrices were considered. The measurements were transformed to the image domain by the transpose operator, $\Phi^T$ before feeding into a CNN based classification network. Only the CNN classifier weights are learned, no learning of the compression matrix is performed. In~\cite{compressedLearning_Elad_2018} an end-to-end compressed learning (CL) framework is studied with respect to image classification. The basic signal model described by (\ref{eq:recon}) and (\ref{eq:syn}).  The compression matrix, $\Phi$, the sparsity basis, $\Psi$, and the classification neural network, represented with the weights $u$, were learnt jointly. Here each of the component is implemented as a linear neural network layer. The reconstruction of the signal was performed by multiplying the measurements, $y$, with the learnt $\tilde{\Psi}$, $\hat{x} = \tilde{\Phi}\tilde{\Psi}y$. Thereafter, $\hat{x}$ is input to the classification network to obtained the final classifications. The following optimization was solved,
\begin{argmini}{\Phi, u}
{\frac{1}{N} \sum_{i=1}^N\mathcal{L}(F_{\tilde{u}}(\Phi(\Psi x_i)), y_i).}
{\label{eq:opt_elad_cl}}
{ \tilde{\Phi}, \tilde{\Psi}, \tilde{u} = }
\end{argmini} Here the compression matrix, sparsity basis, and the classifier weights are all learned. In this work, the compression matrix $\Phi$ is represented as a categorical distribution, which is the case in classical compressed sensing. A binary categorical distribution is learned. This has two specific advantages, if used as a sensing matrix one can include acquisition constraints directly into the sensing matrix, e.g. in single pixel imaging, and if used for post-processing one can design compression matrices according hardware constraints and provide more structure to the compression matrix. The GLODISMO~\cite{glodismo} framework is extended to create a compression-recovery-classification pipeline. The next section describes the proposed model.
\label{sec:glocal}
\begin{figure}[!t]
\centering
\includegraphics[width = \columnwidth]{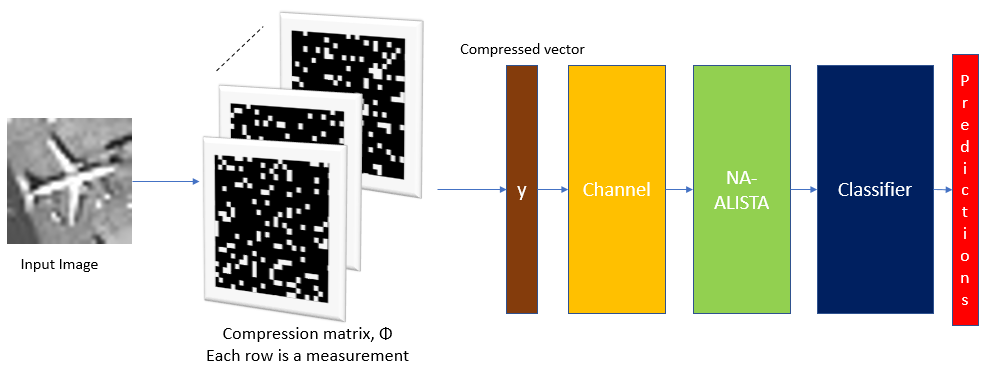}
\caption{GloCal Model.}
\label{fig:glocal}
\end{figure}

\section{GLODISMO + Classifier}

The proposed GloCal (GLODISMO + Classifier) model consists of three parts, as shown in Fig.~\ref{fig:glocal} the compression matrix, $\Phi \in \mathbb{R}^{m \times N}$, a recovery network, $f: \mathbb{R}^m \rightarrow \mathbb{R}^N$ that contains the sparsity basis, and a classifier, $F$. The respective learnable parameters of the components are, $\phi$, $\theta$, and $u$. All the three parts are learned in an end-to-end manner via backpropagation. The compression matrix and the recovery model together form the GLODISMO~\cite{glodismo} sub-network, which is followed by a classifier sub-network, and together they form the GloCal model. 

GLODISMO combines learning of the unrolled reconstruction algorithm and the compression matrix through the following loss function,
\begin{argmini}{\Phi, \theta}
{\mathbb{E}_{x,w}\left[\mathcal{L}(f_\theta(\Phi x + w), x)\right].}
{\label{eq:glodismo}}
{ \tilde{\Phi}, \tilde{\theta} = }
\end{argmini} It is possible to introduce discrete structural constraints on $\Phi$ by creating partitions $\mathcal{P}(\mathcal{I}) = \{\mathcal{I}_1, \dots ,\mathcal{I}_l\}$ of the index set, $\mathcal{I} := \{1, \dots ,m\} \times \{1, \dots, N\}$. $\Phi$ follows a joint distribution over random variables supported on $\mathcal{P}(\mathcal{I})$s. For example one can create partitions of each row and assign a fixed no. of ones to each row. As each row represents a binary linear filter (or a measurement) this creates $m$ sampling masks. This is the structural constraint that is used in this work. However, it is not possible to propagate gradients through a  such a categorical distribution. In~\cite{glodismo}, each partition, $\mathcal{P}(\mathcal{I}_i)$ is independently obtained from a Gumbel reparameterization. During the forward pass IID Gumbel noise is added to each partition elementwise and the indices corresponding to top $d_i$ values are selected, $d_i$ representing the no. of ones in each partition. During the backward pass, instead of the gradient of the hard top-$d_i$ values, the gradient of the softmax with a temperature $\tau$ is used, which makes learning via backpropagation possible. More details can be found in~\cite{glodismo}. The recovery algorithm used is this work is the NA-ALISTA~\cite{naalista} that improves upon previous unrolled algorithms for $l_1$ based recovery by learning the thresholds and step sizes for individual vectors during reconstruction through LSTM networks. A classifier network or model designed for the downstream task follows the NA-ALISTA network. In this work classification is considered as the downstream task. Two forms of loss is considered for end-to-end training. One is the classical cross-entropy loss for the classifier where only the final predictions are considered for training. Second, reconstruction mean squared error (MSE) loss between the NA-ALISTA output and the original image is added to the classification loss. This uses the loss from both the sub-networks for end-to-end learning. It is found that the addition of the recovery loss provides robustness in noisy settings at lower compression ratios.

\section{Results}
\label{sec:results}

This section discusses the datasets, the models, training and testing procedures, and the results for the proposed GloCal model and the benchmarking model. The models were implemented in PyTorch~\cite{pytorch2} on a Intel(R) Xeon(R) W-2245 CPU @ 3.90GHz  Windows machine with 64 GB RAM and an Nvidia RTX A5000 with 24 GB RAM was used for training the models. The MNIST~\cite{mnist} dataset was used to first establish the advantage of the proposed GloCal method. Thereafter, a remote sensing version of MNIST called the Overhead-MNIST~\cite{ovmnist} was used to further explore the performance of GloCal for remote sensing platforms. The datasets are discussed in detail next followed by the models and the experiments. The codes for the presented simulations can be found at~\url{https://github.com/protim1191/glodismo_classifier}.

\begin{figure}[b]
\centering
\includegraphics[width = \columnwidth]{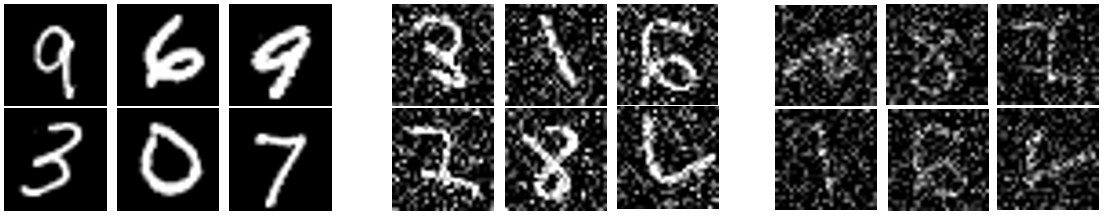}
\caption{Examples from MNIST dataset. \textit{Left column}: Original clean images. \textit{Middle column}: MNIST with AWGN. \textit{Right column}: MNIST with reduced contrast and AWGN. }
\label{fig:mnist-examples}
\end{figure}

\subsection{Datasets}
\label{sec:dataset}
\subsubsection{MNIST}
\label{sec:mnist}
The first dataset used was the MNIST dataset that is a standard benchmark dataset in deep learning. It is a dataset consisting of grayscale handwritten digits. The image size is $28 \times 28$ and there are 60000 samples for training, which were divided into 48000 for training and 12000 for validation. There are additional 10000 samples for testing. To test the GloCal method with noisy input, the n-MNIST~\cite{nmnist} dataset was used, specifically the AWGN (Additive White Gaussian Noise) and Reduced Contrast + AWGN corruptions. The AWGN corruptions are created by adding AWGN to the MNIST with an signal-to-noise (SNR) ratio of 9.5 dB. While the Reduced Contrast + AWGN corruption reduces the MNIST contrast by half and adds AWGN with a SNR of 12 dB. Examples from the clean and noisy MNIST dataset are shown in Fig.~\ref{fig:mnist-examples}.

\begin{figure}
\centering
\includegraphics[width = \columnwidth]{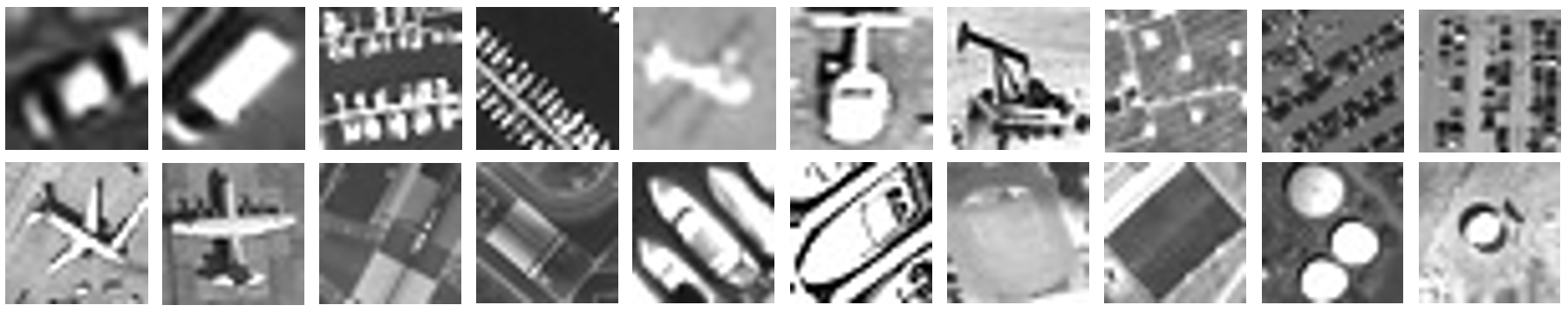}
\caption{Examples from OV-MNIST dataset. The ten classes are car, harbour, helicopter, oil gas field, parking lot, plane, runway mark, ship, stadium, and storage tank.}
\label{fig:ov-mnist-examples}
\end{figure}
\subsubsection{Overhead-MNIST}
\label{sec:ovmnist}

\begin{figure}[b]
\centering
\includegraphics[width = \columnwidth]{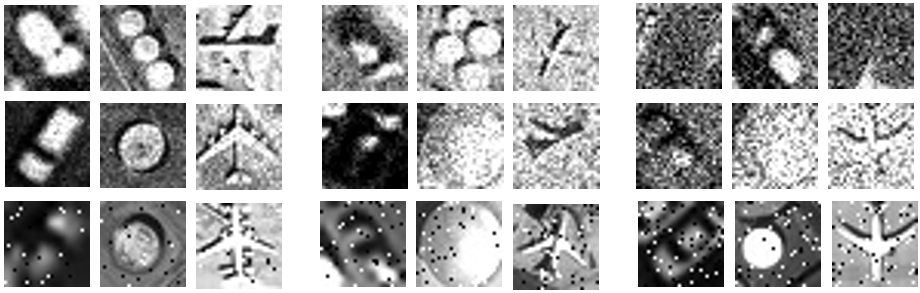}
\caption{Noisy examples from OV-MNIST dataset using noise models from ImageNet-C. \textit{Top panel}: AWGN, \textit{middle panel}: shot noise, and \textit{bottom panel}: impulse noise. \textit{Left column}: severity = 1, \textit{middle column}: severity = 2, and \textit{right column}: severity = 3.}
\label{fig:noisy_examples}
\end{figure}
The Overhead-MNIST~\cite{ovmnist} (OV-MNIST) dataset was used to test the proposed framework in the context of remote sensing images. OV-MNIST is a MNIST like dataset consisting of images from different remote sensing sources. The dataset is curated from sources such as xView~\cite{xview}, UCMerced Landuse~\cite{ucmerced}, DOTA~\cite{dota}, and SpaceNet~\cite{spacenet}. Ten classes each with 1000($\pm 20\%$) images from all the sources were considered. The images were converted to grayscale and cropped to $28 \times 28$. A total of 9584 images were compiled and a $90:10$ training-test split is done. The dataset can be downloaded from~\cite{ovmnist_dataset_gh}. For the experiments the training set is further split into training and validation in $90:10$ ratio. Representative examples from the OV-MNIST dataset are shown in Fig.~\ref{fig:ov-mnist-examples}. A general comment about the OV-MNIST dataset is that even though it is a drop-in replacement for the original MNIST but it more complicated as scenes are diverse and the different resolutions of different remote sensing payloads introduce intrinsic noise that are not found in other MNIST-like datasets. To generate noisy data the~ImageNet-C~\cite{imagenetc} library was used. Three common noise sources present in camera systems is considered, namely AWGN, that represents the thermal noise and dark current noise in the system, photon noise or shot noise due to low light conditions, and impulse noise representing A/D saturation and bad pixels in the camera. Noisy examples are shown in Fig.~\ref{fig:noisy_examples}. It can be seen that with increasing noise the clutter in the image increases obscuring the object of interest increases thus, making the classification task more difficult.

\begin{figure}
\centering
\includegraphics[width = \columnwidth]{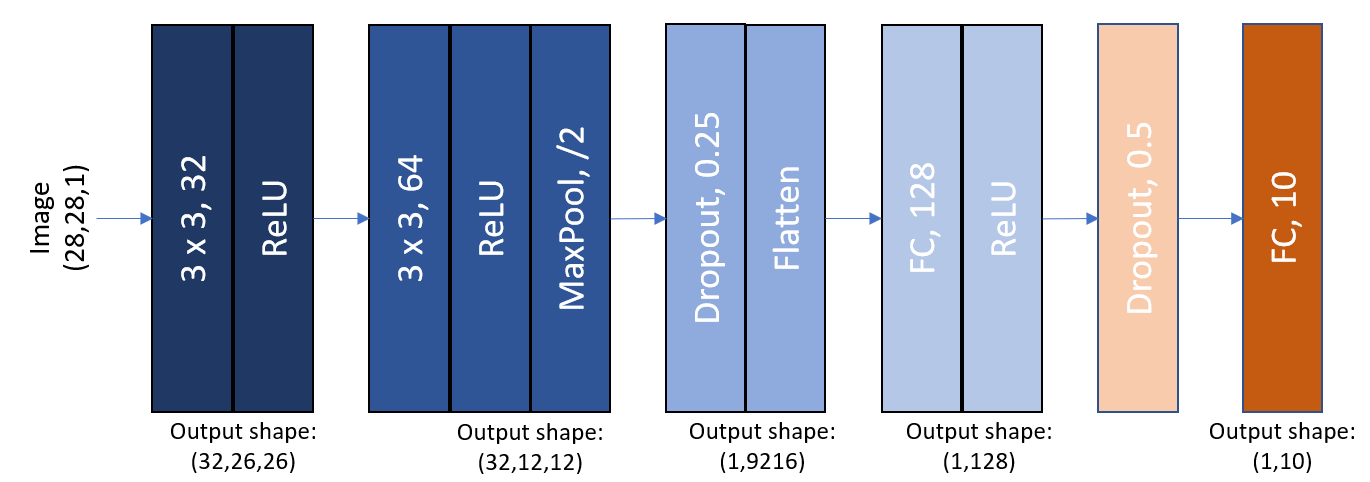}
\caption{Classifier model.}
\label{fig:classifier_model_summary}
\end{figure}

\subsection{Models}
\label{sec:models}
 The model from~\cite{compressedLearning_Elad_2018} is compared with the proposed GloCal network and is referred to as Zisselman-E2E for the rest of the article. The classifier model is the same for all the studied models. The classifier architecture is shown in Fig.~\ref{fig:classifier_model_summary}. It consists of two CNN layers with 32 and 64 filters respectively. The CNNs are followed by ReLU activations. A Maxpool layer follows the second ReLU layer that reduces the spatial size of the CNN feature to $12 \times 12$. This is followed by a dropout layer with a probability of 0.25 and a flatten layer. The flattened tensor is fed to a linear layer with a ReLU activation and is followed by a fully connected linear classifier. A dropout layer with probability of 0.5 is placed before the classifier layer to reduce overfitting. The total no. of trainable parameters is approximately 1.2 million. The compressed sensing portion of the Zisselman-E2E architecture consists of a linear layer with no. of input nodes equal to the number of pixels in the input image and the no. of output nodes corresponds to the chosen compression ratio, $R$. This is followed by a linear reprojection layer that expands the compressed measurement to the original no.of  pixels. A tanh activation is applied followed by another fully connected linear layer. The output vector from the last linear layer is then reshaped to the size of the input image and fed to the classifier. Th GloCal model is explained in Sec.~\ref{sec:glocal}. Two variants are used. One where the loss function is only the classification loss at the end of the complete network, denoted by \emph{GloCal-without-rec-loss} and one where loss consists of both the reconstruction loss from GLODISMO sub-network of GloCal and classification loss from the classifier sub-network of GloCal and is denoted by \emph{GloCal-with-rec-loss} in the following results.

\subsection{Simulations}
\label{sec:simulation}
For the experiments for the classifier baseline and the Zisselman-E2E model mean-variance normalization was performed with respective values of $0.4685$ and $0.2683$ for the OV-MNIST dataset and $0.1307$ and $0.3081$ for the MNIST dataset. The normalization constants are obtained by calculating the mean and variance across all the pixels of the training dataset. For the GloCal model the images were scaled between 0 and 1 by dividing with 255 (original int8 images), were flattened and then input directly to the network. For the OV-MNIST dataset augmentation with rotation of 45° and brightness change of 10\% was implemented for all models.
\begin{table}
\caption{Classification accuracy on the test data of MNIST and n-MNIST for different models and $R = 0.1$. Green denotes the best performing model.} 
\label{tab:acc_mnist}
\begin{center}  
\begin{tabular}{|c|c|c|c|c|}
\hline
\bfseries\thead{Dataset} & \bfseries\thead{Classifier}  & \bfseries\thead{Zisselman-E2E} & \bfseries\thead{GloCal\\(without rec. loss)} & \bfseries\thead{GloCal \\ (with rec. loss)}\\ \hline
     Original MNIST & \cellcolor{green}0.9915  & 0.9827 & 0.9819&0.9840\\ \hline
     n-MNIST AWGN & 0.8484 & 0.9167 & 0.9147 & \cellcolor{green} 0.9639 \\ \hline
     n-MNIST Reduced Contrast AWGN& 0.5599 & 0.8117 & 0.7543&  \cellcolor{green}0.8778\\ \hline
      
\end{tabular}

\end{center}
\end{table}

The classifier baseline and GloCal models were trained with the Adam optimizer with a learning rate of 0.001. The Zisselman-E2E model was trained with SGD as suggested in the authors of~\cite{compressedLearning_Elad_2018} with a learning rate of 0.0025 for MNIST and 0.001 for OV-MNIST and a $l_2$ regularizer weight decay value of 1e-4. Models were trained for a compression ratio of 0.25 and 0.1 for OV-MNIST. Noisy data was only used for testing the models, no noisy data is seen by the models during training. For both the end-to-end networks, Zisselman-E2E and GloCal, the compression-recovery sub-networks and classifier sub-network were pre-trained and the complete network was initialized with the best weights of the sub-networks and retrained end-to-end. For the GloCal-with-recovery-loss a weighted loss was used, $$\mathcal{L} = \alpha\mathcal{L}_{classification} + (1 - \alpha)\mathcal{L}_{recovery}$$ with $\alpha = 0.8$. A grid search was performed for initializing the GLODISMO compression-recovery sub-network in GloCal as described in~\cite{glodismo}. Fixed representation basis was used for the GloCal models, bi-orthogonal 2.2 wavelet with two layers of decomposition. For the NA-ALISTA the no. of unrolled layers was fixed to 16 and the no. of LSTM cells were 64. Sparsity was set to 50, which was calculated from the average wavelet sparsity across the OV-MNIST dataset, for MNIST the sparsity value was set to 20. The sparsity is used within the NA-ALISTA framework to estimate the support of the reconstructed vectors. The discrete structural constraint on $\Phi$ is defined as the no. of ones in each row of the matrix and is fixed to 32 for all experiments, $d = 32$. For OV-MNIST the batch size for Zisselman-E2E was 128 while the for GloCal models was 64. For MNIST, batch size for all models was 128. 

First the results with MNIST with $R = 0.1$ is discussed. The accuracy values for the different models and test data are recorded in Table~\ref{tab:acc_mnist}. The n-MNIST is only used as test data. The baseline classifier works best with the original MNIST, however the Zisselman-E2E and the GloCal models are not far behind. However, when considering noisy data the performance of the simple classifier degrades. The end-to-end network and the GloCal-without-rec-loss perform similarly for n-MNIST AWGN but the Zisselman-E2E outperforms GloCal-without-rec-loss for the n-MNIST Reduced Contrast and AWGN. However, adding a recovery loss to the GloCal provides robustness to the model, as it can be seen to outperform the other models significantly in noisy settings. 

\begin{figure}
\centering
\begin{subfigure}[b]{\columnwidth}
         \centering
         \includegraphics[width=\textwidth]{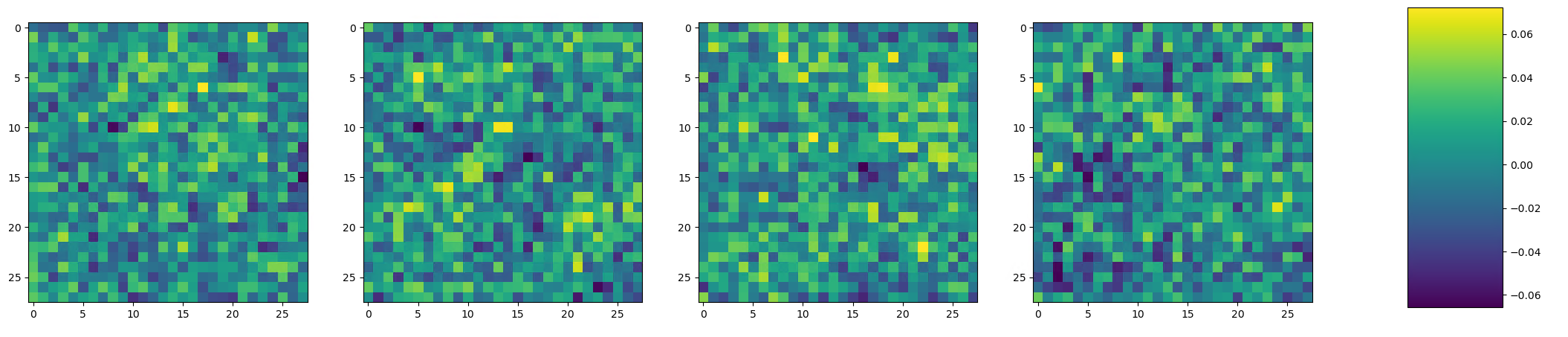}
         \caption{Sensing matrix layer weights from Zisselman-E2E.}
         \label{fig:adler_eg_meas_78}
\end{subfigure}
\begin{subfigure}[b]{\columnwidth}
         \centering
         \includegraphics[width=\textwidth]{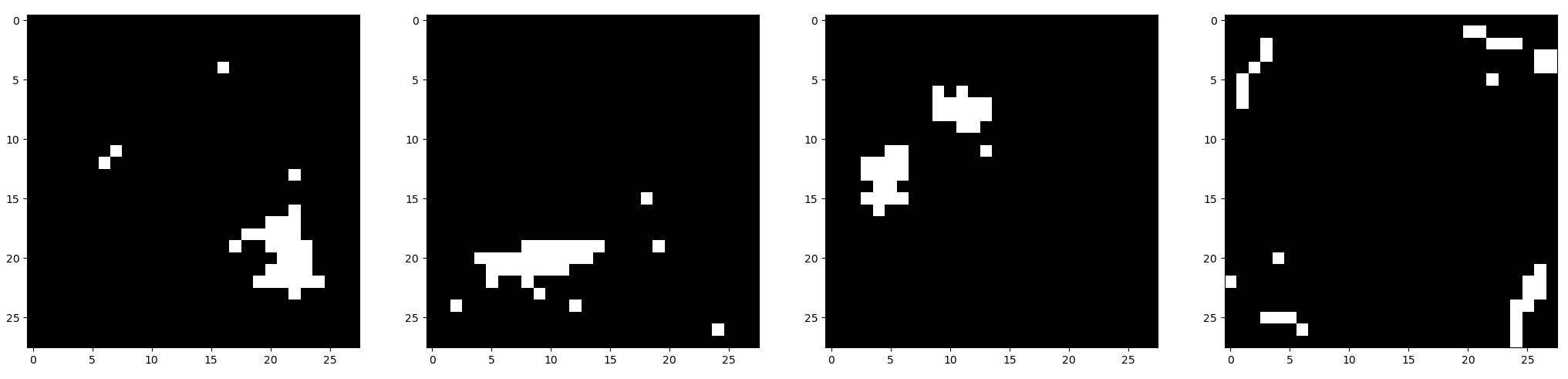}
         \caption{Binary matrix from GloCal with $d = 32$.}
         \label{fig:glocal_eg_meas_78}
\end{subfigure}     
\vspace{+1\baselineskip}
\caption{Example learned measurements for $R = 0.1$. Each image denotes a row of $\Phi$.}
\label{fig:sensing_matrix_examples}
\end{figure}

\begin{figure}\centering
\includegraphics[width = \columnwidth]{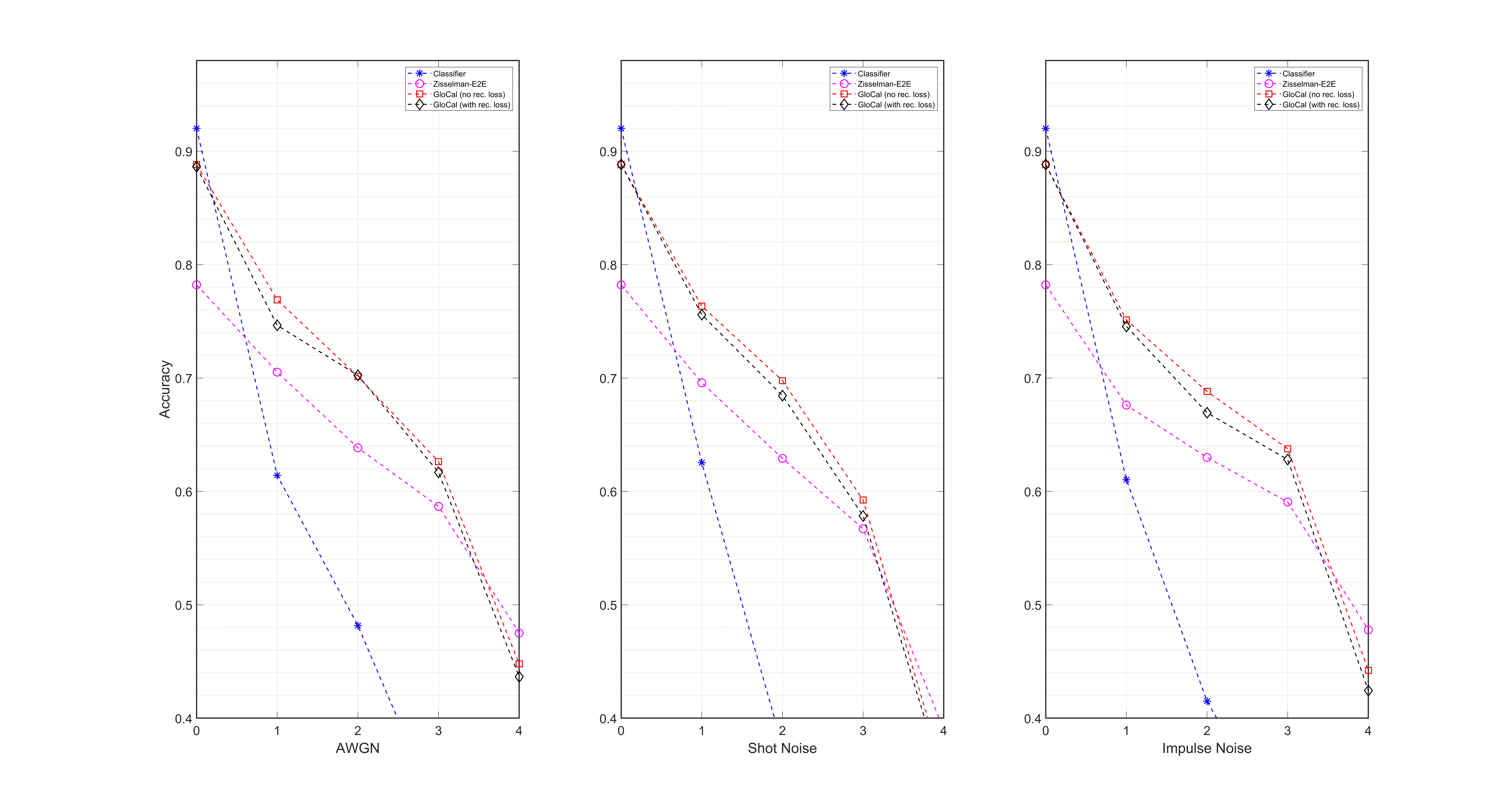}
\caption{Performance of different models with camera noise with $R = 0.25$. The x-axes represent severity according to ImageNet-C.}
\label{fig:camera_noise_perf_m_196}
\end{figure}

\begin{figure}\centering
\includegraphics[width = \columnwidth]{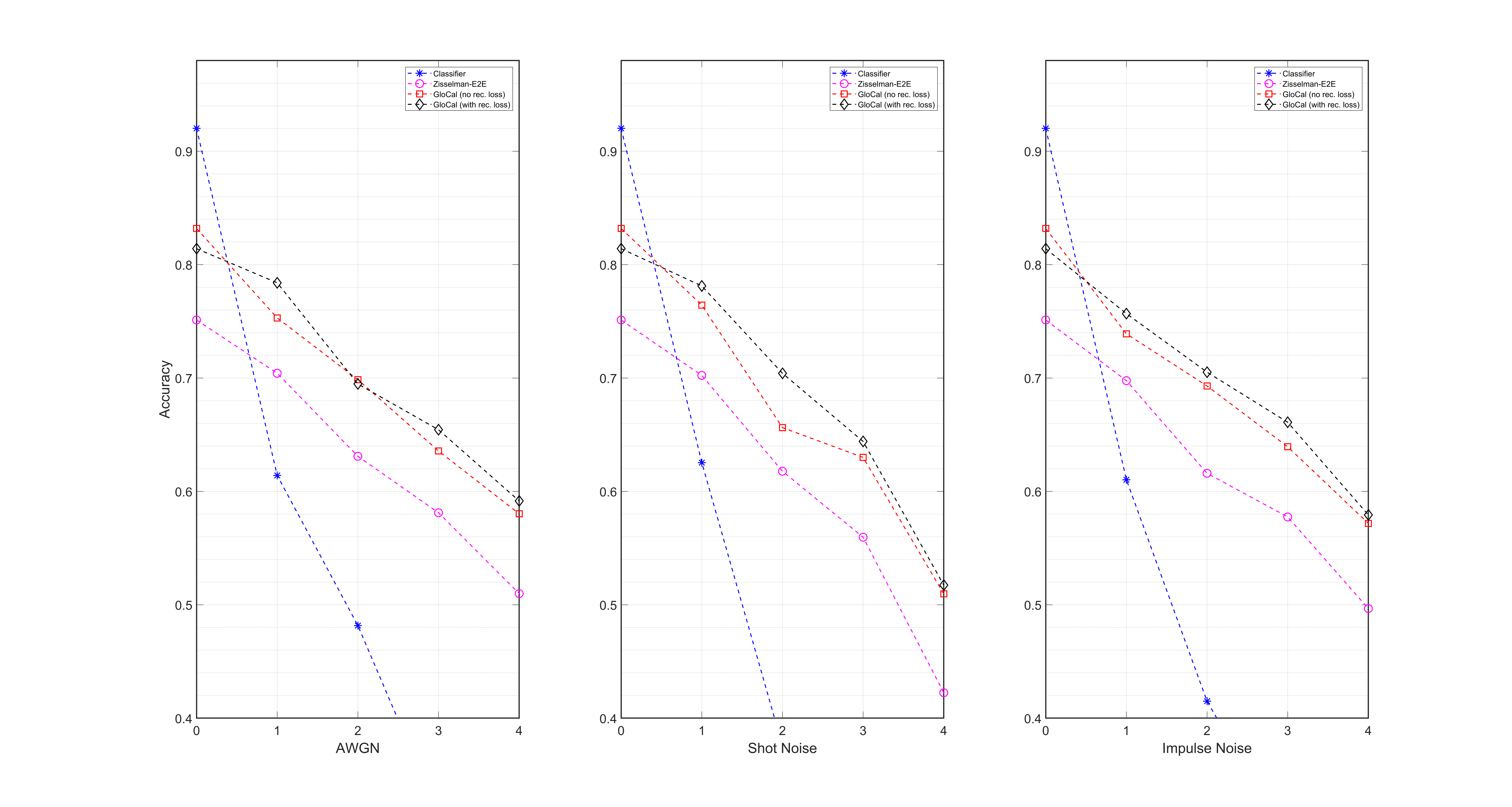}
\caption{Performance of different models with camera noise with $R = 0.1$. The x-axes represent severity according to ImageNet-C.}
\label{fig:camera_noise_perf_m_78}
\end{figure}

\begin{table}[hb]
\caption{Classifier model performance for OV-MNIST (No compression baseline).} 
\label{tab:acc_ov_mnist_classifier}
\begin{center}  
\begin{tabular}{|c|c|c|c|} \hline
& \bfseries\thead{Train} & \bfseries\thead{Val.}& \bfseries\thead{Test} \\ \hline
\makecell{Classifier} & 0.9763  & 0.9354 & 0.9209 \\ \hline
\end{tabular}
\end{center}
\end{table}
Next the models were trained and tested on the OV-MNIST dataset.  Representative examples of learned compressive measurements vectors for Zisselman-E2E and for GloCal is shown in Fig.~\ref{fig:sensing_matrix_examples} for $R = 0.1$. The GloCal measurement vectors are binary in nature as was designed. It can be seen that the learning procedure concentrates on different areas of the image with different measurements thus being able to capture pertinent information from the image required for the downstream task. The no. of ones is a hyperparameter and controlled through the value of $d$. This contrasts with the weight matrix of the linear layer that represents the Zisselman-E2E compression layer where no specific structure in the measurements can be seen. A similar observation was made in~\cite{compressedLearning_Elad_2018} with respect to PCA matrices. The classifier always has access to the complete image for classification and is treated as a no compression baseline. Table~\ref{tab:acc_ov_mnist_classifier} records the performance of the classifier on the training-validation-test splits of the OV-MNIST. The performance of the considered end-to-end compression models is recorded in Table~\ref{tab:acc_ov_mnist}. We see that both the GloCal models outperform the Zisselman-E2E. The performance of the models with different camera noise is presented in Fig.~\ref{fig:camera_noise_perf_m_196} for $R = 0.25$ and in Fig.~\ref{fig:camera_noise_perf_m_78} for $R = 0.1$. For $R = 0.25$ both GloCal models follow each other very closely with the GloCal-without-rec-loss model having an advantage. The GloCal models degrade more gracefully than the Zisselman-E2E model though at higher noise levels the latter has a slight edge.  However, for $R = 0.1$ the GloCal models outperform the Zisselman-E2E. As seen in-case of the MNIST dataset in Table~\ref{tab:acc_mnist}, the GloCal-with-rec-loss performs better at lower compression ratios and can provide more robustness in noisy scenarios. 

\begin{figure}\centering
\includegraphics[width = \columnwidth]{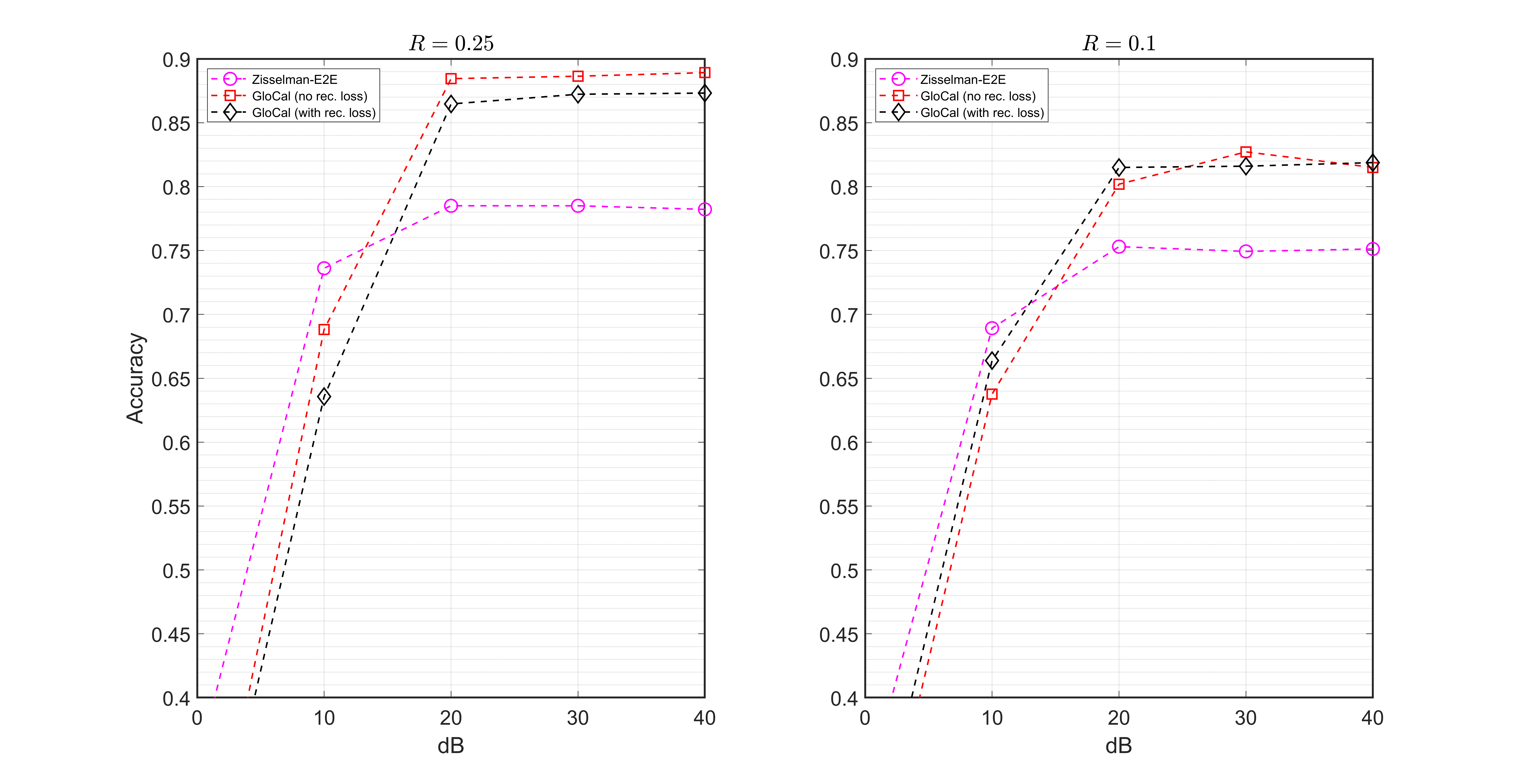}
\caption{Performance of different models with an AWGN channel with OV-MNIST. The x-axes is channel SNR in dB.}
\label{fig:awgn_channel_perf}
\end{figure}

As the proposed GloCal models are to be used for semantic compression in an onboard setting, evaluation with respect to a degrading channel is also preformed. The GloCal framework has a channel noise layer as part of the network, though it was not used for training for degraded channels. For Zisselman-E2E, AWGN was added to the output of the linear compression layer. Thus, for all the models the channel noise was added to the compressed data vector. Figure~\ref{fig:awgn_channel_perf} records the performance of the models with degrading channels for $R = 0.25$ and $R = 0.1$. The GloCal models maintain their noiseless performance at lower noise power. The performance of all the models is near to random chance below 10dB channel SNR. Here as well, the addition of a recovery loss provides better performance at a lower compression ratio.


\begin{table}[!b]
\caption{Classification accuracy for different models and losses for OV-MNIST. } 
\label{tab:acc_ov_mnist}
\begin{center}  
\begin{tabular}{|c|c|c|c|c|c|c|}
\hline
& \multicolumn{3}{c|}{\bfseries\thead{$R = 0.25$}}&\multicolumn{3}{c|}{\bfseries\thead{$R = 0.1$}}  \\ \hline
\bfseries\thead{Model} & \bfseries\thead{Train} & \bfseries\thead{Val.}& \bfseries\thead{Test} & \bfseries\thead{Train} & \bfseries\thead{Val.} & \bfseries\thead{Test}  \\ \hline

Zisselman-E2E & 0.9116 & 0.8226 & 0.7822 & 0.8822& 0.7614 & 0.7512 \\ \hline
\makecell{GloCal \\ (without rec. loss )}& 0.9619 & 0.8978& \cellcolor{green}0.8883 & 0.9021 & 0.8496 & \cellcolor{green}0.8319 \\ \hline
 \makecell{GloCal \\ (with rec. loss )} & 0.9506 & 0.8860  & 0.8864  &0.8888 & 0.8449 & 0.8141\\ \hline
 
\end{tabular}
\end{center}
\end{table}

\section{Discussion and Conclusion}
\label{sec:discussion}
In this work an end-to-end compressed learning framework was designed and studied for application in onboard operations of remote sensing platforms. The GLODISMO~\cite{glodismo} framework was extended to work with downstream tasks; classification was studied as the downstream task in this work.  GLODISMO allows learning of categorical distributions through the Gumbel-softmax trick. This is utilized in this work to learn binary compression matrices that are straightforward to implement in hardware. The compression matrix is learned according to the classification task and thus semantic compression is achieved  through the proposed GloCal (GLODISMO+Classifier) network. The learned matrix can be either used as sensing matrix in single pixel camera applications or can be used as a post-acquisition processing technique on existing hardware for semantic image compression. Two version of the GloCal is proposed. GloCal-without-rec-loss utilizes only the classification loss of the classifier sub-network to learn the complete network while GloCal-with-rec-loss uses both the recovery loss from the GLODISMO sub-network and the classification loss from the classifier sub-network to learn the complete network. The proposed model was evaluated using two datasets, MNIST and OV-MNIST. It was compared with a standard classifier and state-of-the-art model from~\cite{compressedLearning_Elad_2018}. The GloCal was able to outperform the state-of-the-art compressed learning on noisy settings. It was also found that addition of the recovery loss provides robustness to the GloCal model at lower compression ratios. As the models are expected to be deployed onboard remote sensing platforms, performance of the models under a degrading channel was also studied. GloCal models were able to maintain performance against a degrading channel but below 10dB channel SNR, performance of all models reached random chance. However, in GloCal a channel noise layer exists as a part of the architecture and thus one can train the network under different channel noise conditions to introduce robustness in the compressed vector to degrading communications channels. Further, the learned compression matrices in GloCal are binary in nature and can be used as a mask to encode the raw data instead of a performing a matrix-vector product as in~\cite{compressedLearning_Elad_2018}. In the future, the GloCal framework will be extended to RGB and multispectral imaging systems, and also to other downstream tasks such as semantic segmentation.

 

\bibliographystyle{IEEEtran}
\bibliography{IEEEabrv, report} 

\end{document}